 \documentclass[preprint2]{aastex}

\slugcomment{Interferometric Observations of Algol}

\shorttitle{Interferometric Observations of Algol}
\shortauthors{Csizmadia et al.}

\begin{document}


\title{Interferometric observations of the hierarchical triple system 
Algol}


\author{Sz. Csizmadia\altaffilmark{1}}
\affil{Institute of Planetary Research, DLR Rutherfordstr. 2. D-12489, Berlin, Germany \\
       and \\
       MTA Research Group for Physical Geodesy and Geodynamics, 
       H-1585 Budapest, P. O. Box 585., Hungary}
\email{szilard.csizmadia@dlr.de}

\author{T. Borkovits}
\affil{Baja Astronomical Observatory, H-6500 Baja, Szegedi \'ut, Kt. 766, Hungary}

\author{Zs. Paragi}
\affil{Joint Institute for VLBI in Europe, Postbus 2, 7990\,AA Dwingeloo, The Netherlands, }
\affil{and}
\affil{MTA Research Group for Physical Geodesy and Geodynamics,  
       H-1585 Budapest, P. O. Box 585., Hungary}

\author{P. \'Abrah\'am, L. Szabados}
\affil{Konkoly Observatory, H-1525 Budapest, P. O. Box 67., Hungary}

\author{L. Mosoni\altaffilmark{1}}
\affil{Max-Planck-Institut f\"ur Astronomy, K\"onigstuhl 17. D-69117 Heidelberg, Germany}

\author{L. Sturmann, J. Sturmann, C. Farrington, H. A. McAlister, T. A. ten Brummelaar, N. H. Turner}
\affil{Center for High Angular Resolution Astronomy, Georgia State University,
P. O. Box 3969, Atlanta, GA 30302}

\and

\author{P. Klagyivik}
\affil{Department of Astronomy, Roland E\"otv\"os University, H-1517 Budapest,
P. O. Box 32., Hungary}

\altaffiltext{1}{Former address: Konkoly Observatory, H-1525 Budapest, P. O. Box
67. Hungary}

\clearpage

\begin{abstract}
Algol is a triple stellar system consisting of a close semi-detached
binary orbited by a third object.
Due to the disputed spatial orientation of the close pair, the third body
perturbation of this pair is the subject of much research.
In this study, we determine the spatial orientation of the close pair
orbital plane using the CHARA Array, a six-element optical/IR interferometer
located on Mount Wilson, and state-of-the-art e-EVN interferometric
techniques. We find the longitude of the line of nodes for the close pair is
$\Omega_1=48\degr\pm2\degr$ and the mutual inclination of 
the orbital planes of the close and the wide pairs is $95\degr\pm3\degr$.
This latter value differs by $5\degr$ from the formerly known $100\degr$
which would imply a very fast inclination variation of the system, not borne
out by the photometric observations. We also investigated the
dynamics of the system with numerical integration of the equations of
motions using our result as an initial condition. We found large variations 
in the inclination of the close pair (its amplitude $\sim 170\degr$)
with a period of about 20 millenia. This result is in good agreement
with the photometrically observed change of amplitude in Algol's
primary minimum.
\end{abstract}


\section{Introduction}


There are about 1000 triple stellar systems known in the Galaxy, many 
of which consist of a close eclipsing pair and a distant third object orbiting 
around the close pair (Batten 1973, Tokovinin 1997). Algol is probably the 
most well-known of such systems.


Algol consists of a semi-detached eclipsing binary with an orbital
period of 2.87 days (B8V + K2IV) with an F1IV spectral type star
revolving around the binary every 680 days (discovered by radial
velocity measurements, Curtiss 1908). Early interferometric observations
were unable to resolve the system (Merrill 1922), but the third
component was succesfully observed by speckle interferometry (Gezari,
Labeyrie and Stachnik 1972, Blazit et al. 1977, McAlister 1977, 1979)
and its orbit was precisely determined by Bonneau (1979). This result
was refined by using the Mark III optical stellar interferometer (Pan et
al. 1993). 


In the radio regime, Lestrade et al. (1993) detected positional
displacement during the orbital revolution of the AB pair using the VLBI
technique, and identified the K-subgiant as the source of radio
emission. The orbital elements of the close and the wide pairs
determined from all these observations are listed in
Table~\ref{tab:earlierelements}.

The light minima -- mainly primary -- of Algol were extensively observed
in the last two centuries. There was only a very small change in the
eclipse depth during this time. This led S\"oderhjelm (1975, 1980) to
the theoretical conclusion that the mutual inclination of the orbital
planes of the close and the wide pair systems should not be larger than
$11\degr$ and likely they are coplanar because both the shape and depth
of the light minima should have noticeably changed otherwise.  {\bf This
theoretical result was seemingly in good correspondence with the
inclination data deduced both for the close and the wide orbits, i.e.
$i_1=82\fdg3\pm0\fdg2$, and $i_2=83\degr\pm2\degr$ (S\"oderhjelm 1980)
{\bf (see also Figure~ 1)}, respectively, from which S\"oderhjelm (1980)
stated the exact coplanarity\footnote{\bf At this point we should take a
clear distinction between the different kind of orbital elements which
are mentioned in this paper, and describe our notation system. The
optical interferometric measurements give information about the relative
motion of one component to the other. From the CHARA measurements we
get information about the relative orbit of Algol B around Algol A. The
orbital elements refer to this relative orbit denoted by subscript $_1$.
Similarly, the earlier astrometric measurements of the third star give
its relative orbit to the close binary.  (More strictly speaking, to its
photocenter.) These elements are denoted by subscript $_2$. VLBI
measurements give the motion of Algol B component in the sky,
i.e. after the use of the necessary corrections we get the orbital
elements of the secondary's orbit around the center of mass of the
binary. These elements are denoted by subscript $_B$. Finally, in order
to carry out some of the aforementioned corrections for calculating the
orbital motion of Algol B, we need the orbital elements of the close
binary in its revolution around the center of mass of the whole triple
system. These orbital elements are denoted by $_{AB}$ Nevertheless, the
elements of these latter two orbits will be only used when 
necessary.}. Based on the earliest speckle interferometric measurements, 
S\"oderhjelm (1980) calculated $\Omega_2=132\degr\pm2\degr$ for the wide
orbit, and therefore he expected $\Omega_1=132\degr\pm4\degr$ for the 
node of the eclipsing pair. Later Pan et al. (1993) determined the
astrometric orbit of the third component and found that both the longitude 
of the ascending node ($\Omega_2$) and the argument of the periastron
($\omega_2$) of the wide orbit practically differ by $180\degr$ from the
previously accepted values. In the case of an isolated two-body
astrometric orbit, $180\degr$ discrepancy is non-problematic,
because geometrically it means the reflection of the orbital plane onto
the plane of the sky for which transformation the astrometric
coordinates are invariants. Nevertheless, in a triple body system this
results in different spatial configuration of the orbital planes, i.e.
it modifies the mutual inclination fundamentally. 
However, the polarimetric measurements of Rudy (1979) yielded a
contradictory result, suggesting that $\Omega_1=47\degr\pm7\degr$ which
would imply a perpendicular rather than coplanar configuration.
Nevertheless, S\"oderhjelm (1980) suggested, that perhaps the nature of
the polarization mechanism was not understood correctly. Note that Rudy
(1979) resolved also the inclination ambiguity, i.e. he determined that
the angular momentum of the binary directed away from the observer, and,
consequently, the system inclination ($i_1$) should be less than
$90\degr$. }
 

This nearly perpendicular configuration was supported by other
measurements: Lestrade et al. (1993) found $\Omega_1 = 52\degr\pm5\degr$
for the close pair in good agreement with the polarimetric measurements
of Rudy (1979). If this value of $\Omega_1$ is correct, then the mutual
inclination is about $100\degr$ (Kiseleva et al. 1998), and the two
orbital planes are nearly but not exactly perpendicular to each other.
This value for the mutual inclination of the system has been widely
accepted since then. However, {\bf we propose that} this mutual
inclination value cannot be correct due to dynamical considerations 
{\bf which is described hereafter.}

It is well-known that in a hierarchical triple stellar system, the  orbital
planes of the close and  wide pairs are subject to precessional motion,  in
such a way that the normals of the orbital planes move on a conical surface
around the normal of the invariable plane. (Here we omit the effect of 
stellar rotation which is insignificant in an ordinary triple system.)  In
the case of a hierachical triple system where the invariable plane almost
coincides with the wider orbital plane, the precession cone angle is close
to the mutual inclination. Consequently, in the case of the present system,
we would get an almost $160\degr$-amplitude variation in the observable
inclination during the approximate period given by Eq.~(27) of S\"oderhjelm
(1975) (assuming the present approximation is valid, when the mutual
inclination tends to $90\degr$, the precession period tends to infinity). 
Here we refer to Fig.~4 of Borkovits et al. (2004) which clearly shows
that in the case of the aforementioned configuration the observable
inclination of the close binary would have changed by approximately 
$3\degr$ in the last century which evidently contradicts the 
observations. In this case the eclipses would disappear within a few 
centuries. This was already observed in some eclipsing binaries, like in 
SS Lacertae or V907 Sco (see e.g. Eggleton \& Kiseleva-Eggleton 2001) 
but not in Algol.

%

In summary, the polarimetric and the interferometric observations
contradict the coplanar-configuration, but -- since the mutual
inclination is far from exact perpendicularity -- {\bf the latter is} 
not in agreement with the observed tiny change in the minima depth.
A closer approximate perpendicularity of the two orbital planes would
mean that the period of orbital precession becomes so large that the
inclination variation (and consequently the depth variation of the
minimum) of the close pair remains unobservable for a long time,
consistent with the observations.


The aim of this study was to constrain the mutual inclination of the
system better, requiring the measurement of the longitude of the node
for the close pair. The other orbital elements are well-known from
spectroscopic or photometric data, but there is a controversy in the
value of $\Omega_1$. Because the expected apparent size of {\bf the
close binary} semi-major axis is of the order of 2 milliarcseconds
(mas), we carried out optical and radio interferometry measurements. As
we will show, optical and radio interferometry are complementary
techniques. 
{\bf Combining these two, we will show that} it is possible to resolve
the ambiguity in the geometry of the system, and better assess the
accuracy of our measurements.



\section{Observations and data reduction}

\subsection{CHARA Observations}

The CHARA Array is an optical/near-IR interferometer array consisting of six 1m
telescopes. The array is described in detail in ten Brummelaar et al. (2005). 
A detailed overview and further references about the observables and the theory 
of optical interferometry can be found in Haniff (2007).

We observed Algol on three nights (2, 3 and 4 December, 2006) in the
$K_s$ band {\bf (the effective wavelength was $2.133~\mu$m)}. Much of the
second night was lost due to high winds and dusty conditions.


Iota Persei and Theta Persei served as calibration stars. The observations of
the target and the two calibrators were organized into a sequence and the
measurements on calibrators generally bracketed the target observations. We
have 12 data points of Iota Persei, 12 data points of Theta Persei and 23 data
points of Algol itself.

Each data point was calculated from a number of scans. {\bf Each scan
measured} the intensity variations as a function of the path delay.
About 300 scans were collected within five minutes for one
data point, the first 22 of these were obtained on the targets (Algol or
one of the calibrator stars), then 21 scans were done for measuring the
background while the shutter was closed, then more than 200 other scans
were obtained on the targets again and finally 67 further scans for
measuring the background again.

To reduce the data, we used the recipe of McAlister (2002). This
consisted of the following steps: First a low-pass filter was applied to
remove the atmospheric noise. Then the bias was subtracted and the scans
were normalized to unity. As a next step, the scans measured by the two
channels were subtracted from each other (for details, see ten
Brummelaar et al. 2005 and McAlister 2002). This was further processed
by applying a high-frequency filter to reduce noise. Discrete
Fourier-transforms of the scans were calculated and a template was
computed. For this new template, we used the full amplitude for the
frequencies $\pm25/$cycle around maximum frequency and 20\% of the
amplitude for the other frequencies {\bf (for more details see McAlister
2002).} From this we could calculate the maximum deviation of the
template from zero which yielded an estimation of the visibility value.
Removing outlier values, we averaged the remaining values which yielded
the uncalibrated visibility of a particular point. The error was
estimated as the standard deviations of the visibility values of the
more than 200 scans of the point.

The measured visibilities of the calibrators were linearly interpolated for
the times of the Algol observations. A comparison of the true and measured
visibilities of the calibrators yielded a factor which converted the
measured target visibilities to true ones. The true visibilities of the 
two calibrators were estimated as follows:

The uniform disc (UD) angular diameter of Iota Persei is $1.21\pm0.06$
mas according to the '{\it Catalog of High Angular Resolution
Measurements}' (Richichi et al. 2005). The true diameter of the other
calibrator star Theta Persei is $1.201\pm0.015 R_\odot$ which was
determined by the fit of its spectrum (Valenti and Fisher 2005). Since
its HIPPARCOS parallax is known one can easily calculate its true
UD angular diameter to be $0.995\pm0.01$ mas.

The limb-darkened (LD) angular diameter is larger than the UD
diameter. There exists a simple relationship between them (Hanbury Brown 
et al. 1974):
\begin{equation}
\frac{\theta_\mathrm{LD}}{\theta_\mathrm{UD}} = \sqrt{\frac{1 - u_\lambda/3}{1 - 7u_\lambda/15}}
\label{eq:theta}
\end{equation}
The limb-darkening coefficients {\bf for both components} were taken
from the tables of van Hamme (1993). These coefficients are a function
of surface gravity and effective temperature which themselves were 
estimated from the known spectral type of the calibrators. Then
Eq.~(\ref{eq:theta}) yielded the corrections which increase the UD
angular diameters by a few percent only. These corrected values were
used to calibrate the visibilities.

The telescope combinations, epoch of observations, baselines,
uncalibrated visibilities and their errors can be found in
Table~\ref{tab:logCHARA}.  The UV-coverage can be seen in
Fig.~\ref{fig:UVCoverage}

\subsection{e-VLBI Observations}

We observed Algol with a subset of the European VLBI Network (EVN) 
{\bf between 16:37 and 1:19 UT} on 14-15 December 2006 at 5~GHz.
These observations were carried out using the e-VLBI technique, where  the
telescopes stream the data to the central data processor (JIVE, Dwingeloo,
the Netherlands) instead of recording. Using e-VLBI for the observations 
was not fundamental for our measurements, but we took the opportunity of 
an {\bf advertised e-VLBI run} close in time to the CHARA observations, outside the normal 
EVN observing session. The participating telescopes were Cambridge and 
Jodrell Bank (UK),  Medicina (Italy), Onsala (Sweden), Toru\'n (Poland) and 
the Westerbork phased array (the Netherlands). The data rate per telescope 
was 256 Mbps, which resulted in 4$\times$8~MHz  subbands in both LCP and 
RCP polarizations using 2-bit sampling. The
correlation  averaging time was 2 seconds, and we used 32 delay steps (lags). 
Initial clock searching was carried out before the experiment using the 
fringe-finder source 3C345. Algol was
phase-referenced (Beasley \& Conway 1995) to 0309+411 in 3--5--3 minute cycles.  Additional
scans were scheduled on 3C84 for real-time fringe monitoring, and for  
D-term calibration. We used 3C138 to calibrate the Westerbork synthesis 
array amplitudes and polarization.


Post-processing was done using the US National Radio Astronomy Observatory (NRAO) AIPS package
(Diamond 1995). The amplitudes were calibrated using the known antenna gaincurves and  the
measured system temperatures. The data were fringe-fitted, bandpass calibrated, and then
polarization calibrated. We corrected for the polarization  leakage D-terms and fringe-fitted
the cross-hand data, after which the data were averaged in frequency in each subband. Besides
the standard procedure, we used the  WSRT synthesis array measurements on 0309+411 to obtain a
more accurate VLBI flux scale. The phase-reference source showed a low level of circular
polarization (fractional  CP $\sim0.28\%$). The left and right-handed VLBI gains were separately
adjusted in accordance with the WSRT measurement. Imaging was carried out in Difmap (Shepherd
et al. 1994). The snap-shot images (from about 45 minutes data each) were made by
Fourier-transforming the observed visibilities, no self-calibration was applied. 
{\bf We fit circular Gaussian model components to the $uv$-data in Difmap. Initially one component 
was fit in each snapshot. Then, the size and flux of the component was fixed, and we let the
position vary for each 5-minutes Algol scan.}

The log of the observations as well as the calculated individual relative positions can be found 
in Table~\ref{tab:logEVN}.

\section{Data analysis and results}


\subsection{Analysis of CHARA data}


According to the van Cittert-Zernike theorem, the amplitude of the
 visibility is the normalized Fourier-transform of the intensity
distribution (for a comprehensive explanation, see Haniff 2007):
\begin{equation}
V(u, v, t) = \frac{\int \int I(x,y,t)\cos\left(2\pi\frac{u(t)x + v(t)y}{\lambda}\right)\mathrm{d}x\mathrm{d}y}{\int \int I(x,y,t)\mathrm{d}x\mathrm{d}y}
\label{eq: V(uvt)}
\end{equation}
%

%
In this equation $V$ is the true visibility at the ($u$, $v$) spatial  frequencies, ($x$, $y$)
are the corresponding sky coordinates, $t$ is the time, $I$ is the intensity at the ($x,y$) sky
point and finally $\lambda = 2.133~\mu$m is the effective wavelength of our observations
(corresponding to the $K_{s}$ band). Note that the intensity distribution on the sky normally
changes very slowly with time, but we had to include the time dependence in Eq.~(\ref{eq: V(uvt)}) because of the
rapid orbital revolution of the AB pair. Because of the power of CHARA, it is not enough to
model the system as the combination of two point-like sources but we need to combine the
pictures of two extended sources.

To determine this intensity distribution we developed a model which was very close to the one of
Wilson and Devinney (1971) which is based on the Roche-model (Kopal 1978) but it was implemented
in IDL to restore the surface intensities into a matrix and to calculate the sky-projected
picture of the system.


Since we have only about two dozen visibility measurements, we wanted to
limit the number of free parameters, choosing just three: the angular
size of the semi-major axis (in mas), the $K_{s}$ surface brightness
ratio, and the angle $\Omega_1$. {\bf All} other parameters were fixed
according to the values given in Wilson et al. (1972) or in Kim (1989).

The model outlined above was used to fit the data and a grid
search was carried out on the free parameters:
\begin{itemize}

\item[a)]	The surface brigthness ratio was stepped from $0$ to $1.4$ with a 
		step size of 0.05;



\item[b)]	$\Omega_1$ was stepped from $0\degr$ to $180\degr$ with a 
                stepsize of $5\degr$ (note that the visibility 
		amplitude does not change if we rotate the image by $180\degr$, 
                which leaves a $180\degr$ ambiguity in the {\bf ascending node;}
                this ambiguity in $\Omega_1$ can be resolved with VLBI); and
		
\item[c)] 	the angular size of the semi-major axis was stepped from $1.80$ mas to
                $3.60$ mas with a step-size of 0.005 mas (note that 
		the expected size was $2.50$ mas).

\end{itemize}

Nearly 353 000 models were calculated on this grid, and the $\chi^2$ minimum was found. 
Around the minimum, a new search was carried out with a finer grid 
and about 35 000 new models were computed again. Around the minimum, a polynomial
fit yielded the final values and errors. The results are shown in 
Figs.~\ref{fig:Omegamin} and \ref{fig:CHARAFit}.

The best solution we found is reported in Table~\ref{tab:CHARAresults}.

\subsection{Analysis of e-VLBI data}

The EVN measurements were carried out 
on 14/15 December 2006 {\bf for almost 9 hours}. As the orbital period of the 
eclipsing binary is $P=2.8673$ days, the orbital arc covered was near 13\%. 
During this period a secondary
minimum occurred which was simultaneously observed photometrically by the 50-cm
telescope at Piszk\'estet\H o Station of the Konkoly Observatory, Hungary (see
B\'\i r\'o et al., 2007). One of our goals was to observe possible 
partial occultation of the radio source by the primary component and measure the change 
in the circular polarization properties of the source accordingly. Because
Algol flared during the observations {\bf (see Fig.~\ref{fig:flare})}, this goal could not be fulfilled and will
not be discussed further here.

In contrast to the measurements of Lestrade et al. (1993) who observed 
the Algol on four different nights in near quadrature
phases (i.e. around $\phi=0.25$ and $\phi=0.75$), 
our observations covered a small part of the orbit around  $\phi\sim0.5$, when the arc
projected onto the plane of the sky is the largest. (Naturally the case is the same at
$\phi\sim0$). The advantage of this approach is that the motion of the target can be detected
in a few hours, and the measured positions are only slightly affected by the orbital motion of
the AB-C pair, unlike the case when the data are taken at different epochs. Because of the
short observing time interval, the orbit of the AB pair itself is not well constrained.
{\bf Nevertheless, using a priori known values for most of the other orbital 
parameters, one expects to find a relatively accurate value for the 
longitude of the ascending node.} 

There are two limitations that must be mentioned here, 1) short-term tropospheric and ionospheric
phase fluctuations which cannot be modelled well and limit the astrometric accuracy in short
VLBI-measurements, and 2) the variable structure of the radio source (Mutel et al. 1998) -- the
radio emission is not coming from the {\bf surface of the} K-subgiant, but likely from its active polar {\bf coronal} region.
Although Algol was unresolved with our array configuration, the source
centroid position could have changed appreciably because of the bright flare during the run.
These are the factors that must be taken into account in the interpretation of the final result.


We calculated astrometric orbits as well as a simple linear fit on two different sets of
observing data. First, hourly normal points were formed. Then we also calculated the orbit using
5 minutes averages. The latter showed that some points with extremely 
large scatter can lead to
false result in the hourly normal points. Removing such outliers, we 
calculated our final
solution from the 5-minute-average data. Despite the shortness of our observing session (less
than 9 hours) the positional data were corrected for the annual parallax, proper motion and the
revolution around the centre of mass of the triple system. These corrections resulted in
approximately $1\degr$ difference in the node position. For this latter correction we
recalculated the third body orbit by the same code which was used for the binary orbit
determination  from our e-VLBI data. We used both the data sets of Bonneau (1979) and Pan et al.
(1993). From these (very similar results) we applied the orbital elements obtained from Pan's
data (see Table~\ref{tab:kettospalya}) for the wide orbit correction. For the {\bf astrometric} calculations, we
used our own differential correction code based on a Levenberg-Marquardt algorithm, tested
against the data and results of Eichhorn and Xu (1990), and found it to be in excellent
agreement. {\bf In our code the maximum number of adjustable free parameters is nine:} 
the position of the centre of mass ($X_0$, $Y_0$), the orbital period ($P$), and 
the six usual orbital elements ($a$, $e$, $i$, $\omega$, $\Omega$, and $M_0$).
{\bf In the case of the close binary orbit determination, b}ecause of the circular orbit, 
instead of the argument of periastron ($\omega$), and the mean anomaly ($M$), 
their sum, i.e., the true longitude (the distance from the ascending node 
in case of circular orbits) should be used. This was done in 
such a way, that $\omega_1$ was formally considered as zero, while $(M_0)_1=90\degr$ was set for the
mid-(secondary) eclipse moment, $t_0=2\,454\,084.360$. The period and 
the inclination were
acquired from Kim (1989), while the semi-major axis of the secondary's orbit around the centre
of mass of the binary ($a_B$) was calculated from Kim's data. {\bf First we adjusted three parameters
($X_0$, $Y_0$, $\Omega_1$), leaving the other six parameters fixed. 
Finally, the semi-major axis ($a_B$) was also adjusted, as a fourth parameter}.

Using Gnuplot, we also performed a simple linear fit to the data from the knowledge that in the
vicinity of $\phi=0.5$ the motion can be approximated well with a line whose slope gives $\tan\Omega_1$.

Our solutions can be seen in Table~\ref{tab:kettospalya}, as well as in
Fig.~\ref{fig:kettospalya}. 

\section{Discussion}

\subsection{Discussion of the results}

First, we concentrate on the CHARA results. The true size of the semi-major
axis of Algol is $a_1=14.1~R_\odot$ (Kim 1989)  while its angular size was
measured by us to be $a_1=2.28\pm0.02$ mas (see Table~\ref{tab:CHARAresults}). 
By dividing  these two numbers one can find that the distance to Algol 
is $28.6\pm0.3$ parsec. The HIPPARCOS parallax yielded $28.46^{+0.75}_{-0.71}$ parsec. 
The agreement is excellent.

%

The determined surface brightness ratio (0.33) can be converted to a
luminosity ratio by multiplying with the ratio of surface area of the
two stars, which is known from the light curve solution (Wilson et al.
1972; Kim 1989). The resulting luminosity ratio is 0.43. This value is
very close to the photometrically estimated 0.44 (Murad and Budding
1984).


According to the CHARA results, the longitude of the node is $\Omega_1=48\degr\pm2\degr$ 
with an ambiguity of 180 degrees. Because the
determined distance and luminosity ratios agree very well with earlier
measurements obtained with other methods, we have confidence in our
results. With the VLBI measurements (see below) we resolve the
$\pm180\degr$ ambiguity and conclude that $\Omega_1 = 48\degr\pm2\degr$.
This is in excellent agreement with the value determined from
polarimetric measurements ($\Omega_1 = 47\degr\pm7\degr$, Rudy 1979),
indicating that polarimetry is an efficient tool to determine the
spatial orientation of the orbits.

At this point we can determine the mutual inclination $i_m$  with the 
following formula:
\begin{equation}
\cos i_m = \cos i_1 \cos i_2 + \sin i_1 \sin i_2 \cos(\Omega_1 - \Omega_2)
\end{equation}
The result is $i_m=95\degr\pm3\degr$ (the uncertainty reflects the
uncertainties not only in $\Omega_1$ but also in the other angular 
elements), confirming the conclusion of Lestrade et al. (1993) that the two orbital
planes are nearly perpendicular to each other.  This value is, however,
closer to the exact perpendicularity than the $100\degr$ given in Kiseleva et
al. (1998) which was based on the measurements of Lestrade et al. (1993).
Nevertheless, the exact perpendicularity is within the three sigma range. 

Regarding the e-VLBI measurements (see Table~\ref{tab:kettospalya}), 
{\bf we obtained $\Omega_1=53\degr\pm826\degr$ and 
$\Omega_1=53\degr\pm239\degr$ for the
node from the three and four adjusted parameter astrometric fits respectively,} 
and $\Omega_1=52\degr\pm3\degr$ from the simple linear {\bf LSQ fitting}. These
values are close to those obtained previously. However, we 
note that in the case of the
astrometric orbit fittings, the formal errors are extremely large. This naturally reflects the
fact that our measurements cover only a very short fraction of the orbit, and especially in
that phase, where the expected astrometric orbit is almost a straight line. Consequently,
without any a priori information, the orbit would be completely {\bf undeterminable}. However, in this
particular case, the longitude of the node itself is very well determined during this phase, as
this is nothing other than the slope of the obtained straight line. This is well represented by
our linear fit which gives only a minor formal error. So, we think that despite the large formal
errors of the astrometric fittings, the obtained $\Omega_1$ value, at 
least {\bf for the case in hand} should be correct.


{\bf We have to remark that the displacement of the radio source during our 
observing session was almost twice the value which was expected from the 
pure orbital motion.
Formally, of course, we were able to fit an astrometric orbit with a
semi-major axis of $a_B=0\farcs0035$, but the semi-major axis
of the secondary's absolute orbit should be $a_B=0\farcs0019$. Nevertheless, although our 
four-adjustable-parameter fit gave an unreastically large value for the semi-major axis,
and consequently, should be rejected, it gave the same value for 
$\Omega_1$ as
the three-parameter (fixed $a_B$) fit. This also suggests, that despite the large
formal errors, the value obtained for $\Omega_1$ seems to be 
well-determined.}
{\bf This apparent large displacement or scatter is likely} the consequence 
of the positional errors caused by short-term atmospheric phase fluctuations, {\bf and} the 
variable structure of the source during the flare. {\bf This would not be unprecedented. Large 
positional change was observed in the RS CVn system IM Peg during a flare by Lebach et al. 
(1999).}
These structural variations and the origin of large radio flares in Algol could be studied
with VLBI array configurations and observing frequencies providing (sub-)mas angular
resolution.

\subsection{Comparison to former VLBI measurements}

Before further discussion of the dynamical consequence of our result, we 
feel it 
necessary to comment on the well-known VLBI result of Lestrade et al. 
(1993).
In our opinion it is without 
question that the excellent paper of Lestrade et al. (1993) is epoch-making in its
significance, but, unfortunately, at the last step of their analysis they made some mistakes. As we cited earlier they obtained $\Omega_1=52\degr$. A
careful look at their Fig.~3 clearly shows that this cannot be the 
correct result. One can
see in that figure, that the coordinate difference is larger in the declination direction than
in the right ascension one. Consequently, the slope of the straight line fitted to their four
points should be less than $45\degr$, at least, when it is measured from north to east (i.e.
from $\delta$ to $\alpha$). So, in our opinion, they obtained their 
value by measuring from east to
north, and so their correct result should be $\Omega_1=38\degr\pm5\degr$. {\bf Furthermore, 
we found, that the exchange of the ($\delta$, $\alpha$) coordinate pairs was not limited 
only to the determination of $\Omega_1$, but it was applied in their all astrometric 
calculations. A less critical further consequence is that they obtained 
a reversed orbital revolution
(compare their Fig.~4 with our Fig.~\ref{fig:Lestrade}). However, the 
case
of the correction for the orbital motion in the triple system is more problematic.
Due to the aforementioned exchange of coordinate pairs, the direction of 
the orbital revolution in the 
wide orbit is also reversed, and, consequently, the correction of the four observed coordinates 
for the orbital motion in the triple system is erroneous. }

Fig.~\ref{fig:Lestrade} shows  the corrected data points together with 
Lestrade et al.'s original solution.
{\bf As one can see} we obtain somewhat larger scatter in the data 
points. 
For these points we obtained $\Omega_1=45\degr\pm20\degr$ 
{\bf from the linear fit. (In this case we do not calculate an 
astrometric fit,
as practically only two data points are known for the orbit. 
Remember, point one and two, as well as three and four belong almost to 
the same orbital phase, respectively.)}

{\bf We should note that in the case of the simple linear fits, 
the probable errors say nothing about the physical reliability of 
the results, or the accuracy of the measurements. They simply indicate  
the possibility to fit one simple line for the four data points. To clarify 
this statement we have to keep in mind that the four points practically 
belong to two orbital phases. Consequently, theoretically the first two points 
should practically coincide, and the same is also true for the third and fourth
ones. Instead of this, one can see that the distances of points one 
and two are $\approx 0.9$ mas and $\approx 1.3$ mas according to Lestrade's 
and our corrections, respectively, while for the other two points these values 
are $\approx 1.0$ mas and $\approx 1.8$ mas, respectively. These  distances 
seem to be in good agreement with the statement of Mutel et al. (1998) about 
the radio source of Algol B, i.e. ,,The structure is double lobed with a 
separation of $1.6\pm0.2$ mas (1.4 times the K star diameter)'' 
So this means that due to the extended, and presumably varying structure of 
the radio source, we cannot expect larger accuracy from the VLBI position 
measurement. Returning to the question of the probable errors, on  
using the Lestrade et al. (1993) original correction, the four data 
points then  
coincide almost in one straight line, so we can get a better linear fit 
than with our correction but as theoretically we should get only 
two points instead of four, this fact does not give any information about 
the reliability of the two results.
Furthermore, Mutel et al. (1998) conclude that the individual lobes are in 
the polar region, which is in better correspondance with the position of 
the radio source with respect to the orbit, in our ,,less accurate'' solution 
(see again  Fig.~\ref{fig:Lestrade}).}

{\bf Taking into account the large scatter in the positions, 
this is in a very good agreement with the polarimetric
measurements ($\Omega_1=47\degr\pm7$, Rudy 1979) and with our CHARA 
measurements ($\Omega_1=48\degr\pm3\degr$) as well. }
%

\subsection{Dynamics of the system}

In order to investigate the dynamical behaviour of Algol in the near past and
future, we carried out numerical integration of the orbits for the triple 
system. Detailed description of our code can be found in Borkovits et 
al. (2004). This
code simultaneously integrates the equations of the orbital motions and the
Eulerian equations of stellar rotation. The code also includes stellar dissipation,
but the short time interval of the data allows that term to be ignored. Our input 
parameters for Algol AB were
almost identical with that of Table~\ref{tab:earlierelements} with the
exception of $\Omega_1$ which was set to $48\degr$ in accordance with 
our CHARA result. The orbital elements of the wide orbit were taken from
Table~\ref{tab:kettospalya} {\bf (with two natural 
modifications, namely, instead of $a_{AB}$ and $\omega_{AB}$, $a_2$ and $\omega_2$ were used)}. As
further input parameters, the $k_2$, $k_3$ internal structure constants for
the binary members were taken from the tables of Claret and Gimenez (1992) as
$k_2^{(1)}=0.0038$, $k_3^{(1)}=0.0011$,  $k_2^{(2)}=0.0240$,
$k_3^{(2)}=0.0087$, respectively. 

Our numerical results between 1600 and 2100 AD can be seen in
Fig.~\ref{fig:Algol_500years}. The variation of inclination between 7500 BC
and 22\,500 AD was also computed and can be seen in Fig.~\ref{fig:Algol_long}.
Note that Algol AB does not show eclipses when the inclination is lower than 
$63\degr$ or higher than $117\degr$. It shows partial eclipses if the 
inclination is between $63\degr$--$117\degr$ and moreover, it shows total eclipses when the
inclination is between $87.3\degr$--$92.7$. Therefore the last time when Algol was not an
eclipsing binary was before 161 AD and it showed partial eclipses between 161 AD
and 1482 AD with increasing amplitude. Of course, at the beginning of this
period, the eclipses featured a very small amplitude which later 
increased. By 1482 AD, the eclipses became total, and this was the case
until 1768 AD. The maximum length of the totality was about 0.5 hours around
1625 AD. It is wortnoting that in Algol the brighter star is the smaller
one. Therefore the darker component could totally cover the brighter object
causing large depth of minima of 2.8 magnitudes, so for a naked-eye 
observer it would almost disappear from the night for half an hour since its 
brightness during
this half hour would be about 5.0 magnitudes. (The amplitude nowadays is 
only about
1.3 magnitudes). {\bf Note, that during the totality the light of the wide, 
C component is the dominant.} As one can see in these diagrams in the time of the discovery
as a variable star (Montanari 1671), the inclination of the close pair was about
$88\degr$, making discovery easier.
Our results also suggest that the light variation of 
Algol might have been known in the medieval Arabic and Chinese civilisations
 (e.g. Wilk 1996). However, it should be emphasized that this time-data 
are rough
approximations only. Since we could not determine the position of the
node better than $\pm2\degr$, and for exact calculations one needs an 
accuracy better by one {\bf order of magnitude}, these 
numbers should be refined in the future.

From about 1768 AD Algol show partial eclipses until approximately
3044 AD and the depth of the minima decreased in good agreement with the 20th
century photometry measurements (S\"oderhjelm 1980). 
Considering the scientific era, one can see that our result suggests an
inclination variation of $\Delta i\approx-1\fdg6$ in the last century which
is in accordance with the statement of S\"oderhjelm (1980).

%

\section{Conclusions}

In this study we focused on the orientation of orbital planes in the
hierarchical triple stellar system Algol. This is an important issue because
the system has been showing a stable eclipse light curve for centuries: this
could happen if the orbital planes of the close pair and of the third body are
either almost coplanar or perpendicular to each other (S\"oderhjelm 1980,
Borkovits et al. 2004). However, former estimations (Kiseleva et al. 1998
based on the results of Lestrade et al. 1993) showed that the mutual
inclination is $100\degr$  which would yield a fast inclination variation and
consequently would result in the disappearance of the eclipses 
(S\"oderhjelm 1975, 1980; Borkovits et al. 2004).


We found that joint use of optical and radio interferometry techniques
in our project had great benefits. While in optical interferometry one cannot 
measure the visibility phase, there is a well understood a-priori source model 
(two stars orbiting each other). Because of this, even with the limited number 
of baselines available, we could fit well the value of the ascending 
node using 
visibility amplitudes only. In the radio regime, we are able to measure both 
visibility amplitude and phase, but we detect only the active corona of one of 
the stars. Unfortunately, this corona is highly variable and the Earth's atmosphere adds 
phase fluctuations that limit astrometric precision in short measurements.
However, {\bf with the combination of the VLBI measurements and the orbital phase information
arising from the eclipses} one can resolve the $\pm 180$ phase ambiguity (once it
is known which component emits in the radio).

After careful analysis, we found the mutual inclination angle of the orbital planes of the close
and the wide pairs to be $95\degr\pm3\degr$. Using this value as an initial value we integrated
the equation of motion of the system back to $-$7500 and forward to $+$22500. This helped to
give support to the notion that medieval civilizations could observe the 
big changes (up to 2.8 magnitudes) of
Algol in the 17th century (Wilk 1996). The rate of inclination change of the close pair was
found to be $\Delta i\approx-1\fdg6$/century in the 20th century which shows only minor
observable changes in the depth and shape of the minima in accordance with the photometric
observations (S\"oderhjelm 1980). Therefore, the regular and precise observations of Algol's
minima are recommended to further refine the geometrical configuration and to better understand
the dynamics of triple stellar systems.



\acknowledgments

{\it Acknowledgements.} The CHARA Array is funded by the National Science Foundation through NSF
grants AST-0307562 and AST-06006958 and by Georgia State University through the College of Arts
and Sciences and the Office of the Vice President for Research. 

e-VLBI developments in Europe are supported by the EC DG-INFSO  funded
Communication Network Developments project 'EXPReS', Contract No. 02662
(http:/$\!$/www.expres-eu.org/). The European VLBI Network (http: /$\!/$www.evlbi.org/)
is a joint facility of European, Chinese, South African and other radio 
astronomy institutes funded by their national research councils.
{\bf ZP acknowledges support from the Hungarian Scientific Research Fund
(OTKA, grant no. K72515).}

This research has made use of the SIMBAD database, operated at CDS, Strasbourg,
France and has made use of NASA's Astrophysics Data System.

We thank the Konkoly Observatory of the Hungarian Academy of Sciences  for the
availability of the 50 cm telescope during the e-VLBI measurements. 

We also thank the kind support of Dr. E. Forg\'acs-Dajka (Roland 
E\"otv\"os University, Budapest, Hungary).



%
%
%
%
%
%
%
%
%

\clearpage

\begin{table*}
\begin{center}
\caption{Orbital elements, and astrophysical parameters of Algol A-B and AB-C 
determined by previous studies.  $a_1$, $e_1$, $i_1$ and $\omega_1$ were taken from
Kim (1989) while $\Omega_1$ is from Rudy (1979).  The elements of the AB-C
system were taken from Pan et al. (1993). The stellar quantities were also
taken from Kim (1989).
\label{tab:earlierelements}}
\begin{tabular}{lcrr}
\tableline\tableline
Quantity & Notation & A-B & AB-C \\
\tableline
Time of periastron (HJD)& $T$      & $2\,445\,739.0030$\tablenotemark{a}& $2\,446\,931.4$  \\
Period                  & $P$      & $2\fd8673285$               & $680\fd05$   \\
Semi-major axis         & $a$      & $0\farcs0023$\tablenotemark{b}& $0\farcs09461$   \\      
                        &          & $14.1 R_\odot$              & $582.9R_\odot$\tablenotemark{c} \\
Eccentricity            & $e$      & 0\tablenotemark{d}          & $0.225$	\\
Inclination             & $i$      & $82\fdg31$                  & $83\fdg98$ \\
Argument of periastron  & $\omega$ & -\tablenotemark{d}          & $310\fdg29$  \\
Longitude of the ascending node & $\Omega$ & $47\degr$           & $312\fdg26$  \\
\tableline
\tableline
Stellar Parameter & Algol A & Algol B & Algol C \\
\tableline
Mass   ($M_\odot$)& 3.8     & 0.82    & 1.8     \\
Radius ($R_\odot$)& 2.88    & 3.54    & 1.7     \\
\tableline

\end{tabular}
\tablenotetext{a}{Time of primary minimum from Kim (1989). Note, if we
set formally $\omega_1=270\degr$ then this gives the time of periastron.}
\tablenotetext{b}{Kim (1989) gave the semi-major axis of the binary in solar
radii ($14.1 R_\odot$). Using the HIPPARCOS parallax we transform it into arcseconds.}
\tablenotetext{c}{Pan et al. (1993) gave the semi-major axis of the third body
in arcseconds. Using the HIPPARCOS parallax we transform it into solar units.}
\tablenotetext{d}{Kim (1989) assumed a circular orbit. Hence eccentricity is
zero and $\omega$ is not defined in {\bf the} circular case. See 
also $^a$.}
\end{center}

\end{table*}

\clearpage

\begin{table*}
\begin{center}
\caption{Log of observations and the observed normalized visibilities of Algol\label{tab:logCHARA}}
\begin{tabular}{crrrrrrr}
\tableline\tableline
 Telescopes& Time (UT)   & u[m] & v[m] & B[m] & V & $\sigma(V)$ \\
\tableline
2006 Dec 2 &          &        &          &         &        &      \\
           &          &        &          &         &        &      \\
W2-S2	   & 05:56:04 & 54.709 & $-$168.117 & 176.794 & 0.763 & 0.031 \\
W2-S2	   & 06:01:56 & 57.224 & $-$167.175 & 176.698 & 0.771 & 0.031 \\
W2-S2	   & 06:36:05 & 71.063 & $-$160.881 & 175.877 & 0.749 & 0.029 \\
W2-S2	   & 06:59:09 & 79.533 & $-$155.896 & 175.011 & 0.802 & 0.030 \\
W2-S2	   & 07:26:00 & 88.366 & $-$149.424 & 173.597 & 0.738 & 0.030 \\
W2-S2	   & 07:48:43 & 94.892 & $-$143.450 & 171.995 & 0.740 & 0.038 \\
E2-W2	   & 09:31:01 & 60.207 &  129.565 & 142.870 & 0.285 & 0.007 \\
E2-W2	   & 09:56:59 & 44.772 &  133.478 & 140.787 & 0.298 & 0.008 \\
E2-W2	   & 10:32:33 & 22.735 &  136.923 & 138.802 & 0.309 & 0.005 \\
E2-W2	   & 10:58:40 &  6.173 &  138.011 & 138.150 & 0.338 & 0.005 \\
E2-W2	   & 11:18:11 & $-$6.268 &  138.009 & 138.151 & 0.419 & 0.003 \\
%
%
\tableline
2006 Dec 3 &          &        &          &         &       &       \\
           &          &        &          &         &       &       \\
E2-W2	   & 03:19:37 & 130.402& $-$0.093   & 130.402 & 0.616 & 0.050 \\
E2-W2	   & 03:26:40 & 132.347&  2.563   & 132.372 & 0.576 & 0.050 \\

\tableline
2006 Dec 4 &          &         &         &         &       &       \\
           &          &         &         &         &       &       \\
E2-S2	   & 05:14:50 & $-$105.730& $-$220.787& 244.798 & 0.455 & 0.050 \\
E2-S2	   & 05:45:49 &  $-$89.340& $-$229.467& 246.246 & 0.539 & 0.050 \\
W1-S2	   & 07:25:08 &  196.777& $-$142.326& 242.854 & 0.601 & 0.020 \\
W1-S2	   & 07:35:15 &  198.818& $-$136.586& 241.215 & 0.711 & 0.026 \\
W1-S2	   & 07:58:24 &  202.018&  123.269& 236.657 & 0.671 & 0.029 \\
W1-S2	   & 08:23:14 &  203.147& $-$108.827& 230.461 & 0.623 & 0.021 \\
W1-S2	   & 08:46:36 &  202.022&  $-$95.239& 223.347 & 0.570 & 0.021 \\
W1-S2	   & 09:12:53 &  198.237&  $-$80.137& 213.823 & 0.560 & 0.041 \\
W1-S2	   & 09:33:28 &  193.441&  $-$68.569& 205.234 & 0.594 & 0.049 \\
W1-S2	   & 09:57:44 &  185.776&  $-$55.361& 193.850 & 0.608 & 0.035 \\
W1-S2	   & 10:19:34 &  177.089&  $-$43.992& 182.472 & 0.636 & 0.004 \\
\tableline
\end{tabular}
\end{center}
\end{table*}

\clearpage

\begin{table*}
\begin{center}
\caption{Log of e-EVN observations and the observed total intensity peak positions of Algol. 
The complete data rows (following the horizontal lines) give the normal points formed
from the each approx. 45-min. long observing scans, while in the  case 
of the five-minutes averages
only the astrometric angular coordinates are given. (See text for details.)
(Algol position used for correlation (nominal coordinates):
RA = $3^\mathrm{h}8^\mathrm{m}10\fs1315$, DEC = $+40\degr57'20\farcs332$ (J2000); [DEC = 40.955648, cos(DEC) = 0.7552172];
Fluxdens: recovered VLBI total flux density during that timerange (mJy);
R: modelfit centroid distance to phase centre (i.e. nominal position); 
$\Theta$: modelfit centroid position angle in degrees, measured from North to East; 
on the map North is top (Y axis) and East is to the left (X axis);
X: measured -- nominal X coordinate in arcseconds ( X$ = \delta\mathrm{RA}\cdot\cos(\mathrm{DEC})$ )
         X [arcsec] = R [mas]$\cdot\sin\Theta / 1000.0$;
$\delta$RA: X transformed to real RA coordinate difference, in seconds;
     $\delta$RA [sec] = X [arcsec]$ / (15\cdot\cos(\mathrm{DEC}))$;
$\delta$DEC: measured -- nominal Y coordinate in arcseconds (=$\delta$DEC);
     $\delta$DEC [arcsec] = R[mas]$\cdot\cos\Theta / 1000.0$) 
\label{tab:logEVN}}
{\small 
\begin{tabular}{crrrrrrr}
\tableline\tableline
 Start (UT)& End (UT) & Fluxdens. & R [mas] & $\Theta$ &    X [$''$]    &  $\delta$RA [s] & Y=$\delta$DEC [$''$] \\
\tableline
  16:37  &  17:18 &  46.2 mJy &  24.62  & 103.4 &   0.02395   & 0.002114 & -0.00571  \\
  16:37  &  16:38 &           &  25.60  & 102.6 &             &          &           \\
  16:41  &  16:46 &           &  24.29  & 103.6 &             &          &           \\
  16:49  &  16:54 &           &  24.86  & 102.9 &             &          &           \\
  16:57  &  17:02 &           &  24.38  & 103.9 &             &          &           \\
  17:05  &  17:10 &           &  24.79  & 103.6 &             &          &           \\
  17:13  &  17:18 &           &  24.56  & 103.0 &             &          &           \\
\tableline
  17:33  &  18:18 &  43.4 mJy &  24.53  & 103.9 &   0.02381   & 0.002102 & -0.00589  \\ 
  17:33  &  17:38 &           &  24.46  & 103.9 &             &          &           \\
  17:41  &  17:46 &           &  24.46  & 103.3 &             &          &           \\
  17:49  &  17:54 &           &  25.15  & 104.1 &             &          &           \\
  17:57  &  18:02 &           &  24.36  & 104.3 &             &          &           \\
  18:05  &  18:10 &           &  24.42  & 103.6 &             &          &           \\
  18:13  &  18:18 &           &  24.32  & 104.2 &             &          &           \\
\tableline
  18:33  &  19:18 &  32.6 mJy &  24.48  & 104.4 &   0.02371   & 0.002093 & -0.00609  \\
  18:33  &  18:38 &           &  24.57  & 104.6 &             &          &           \\
  18:41  &  18:46 &           &  24.54  & 103.9 &             &          &           \\
  18:49  &  18:54 &           &  24.56  & 104.9 &             &          &           \\
  18:57  &  19:02 &           &  24.31  & 104.5 &             &          &           \\
  19:05  &  19:10 &           &  24.42  & 104.3 &             &          &           \\
  19:13  &  19:18 &           &  24.45  & 104.7 &             &          &           \\
\tableline
  19:33  &  20:18 &  25.0 mJy &  24.26  & 105.2 &   0.02341   & 0.002067 & -0.00636  \\
  19:33  &  19:38 &	      &  24.26  & 105.0 &	      & 	 &	     \\
  19:41  &  19:46 &	      &  24.12  & 104.8 &	      & 	 &	     \\
  19:49  &  19:54 &	      &  24.43  & 105.3 &	      & 	 &	     \\
  19:57  &  20:02 &	      &  24.33  & 105.7 &	      & 	 &	     \\
  20:05  &  20:10 &	      &  24.32  & 105.6 &	      & 	 &	     \\
  20:13  &  20:18 &	      &  24.07  & 105.0 &	      & 	 &	     \\
\tableline
  20:33  &  21:18 &  16.5 mJy &  23.81  & 105.5 &   0.02294   & 0.002025 & -0.00636  \\
  20:33  &  20:38 &           &  23.78  & 105.7 &             &          &           \\
  20:41  &  20:44 &           &  24.38  & 104.5 &             &          &           \\
  21:01  &  21:02 &           &  24.02  & 105.3 &             &          &           \\
  21:05  &  21:10 &           &  23.71  & 105.5 &             &          &           \\
  21:13  &  21:18 &           &  23.74  & 105.7 &             &          &           \\
\tableline
  21:33  &  22:18 &  12.6 mJy &  23.78  & 106.2 &   0.02284   & 0.002016 & -0.00663  \\
  21:33  &  21:38 &	      &  23.62  & 106.3 &	      & 	 &	     \\
  21:41  &  21:46 &	      &  24.01  & 105.8 &	      & 	 &	     \\
  21:49  &  21:54 &	      &  23.90  & 106.6 &	      & 	 &	     \\
  21:57  &  22:02 &	      &  23.85  & 106.4 &	      & 	 &	     \\
  22:05  &  22:10 &	      &  23.44  & 106.3 &	      & 	 &	     \\
  22:13  &  22:18 &	      &  23.85  & 105.7 &	      & 	 &	     \\
\tableline
\end{tabular}
}
\end{center}
\end{table*}

\clearpage

\setcounter{table}{2}

\begin{table*}
\begin{center}
\caption{Continuation of Table~\ref{tab:logEVN}}
{\small 
\begin{tabular}{crrrrrrr}
\tableline\tableline
 Start (UT)& End (UT) & Fluxdens. & R [mas] & $\Theta$ &    X [$''$]    &  $\delta$RA [s] & Y=$\delta$DEC [$''$] \\
\tableline

  22:33  &  22:18 &   9.6 mJy &  23.51  & 107.1 &   0.02247   & 0.001984 & -0.00691  \\
  22:33  &  22:38 &	      &  23.57  & 106.7 &	      & 	 &	     \\
  22:41  &  22:46 &	      &  23.38  & 106.8 &	      & 	 &	     \\
  22:49  &  22:54 &	      &  23.30  & 107.6 &	      & 	 &	     \\
  22:57  &  23:02 &	      &  23.67  & 107.5 &	      & 	 &	     \\
  23:05  &  23:10 &	      &  23.36  & 107.6 &	      & 	 &	     \\
  23:13  &  23:18 &	      &  23.75  & 106.9 &	      & 	 &	     \\
\tableline
  23:33  &  00:18 &   8.3 mJy &  22.92  & 107.4 &   0.02187   & 0.001931 & -0.00685  \\
  00:13  &  00:18 &           &  23.20  & 107.2 &             &          &           \\
\tableline
  00:33  &  01:18 &   6.1 mJy &  22.90  & 107.9 &   0.02179   & 0.001924 & -0.00704  \\
  00:33  &  00:38 &	      &  23.19  & 107.7 &	      & 	 &	     \\
  00:41  &  00:46 &	      &  22.58  & 107.6 &	      & 	 &	     \\
  00:49  &  00:54 &	      &  22.73  & 108.9 &	      & 	 &	     \\
  00:57  &  01:02 &	      &  23.03  & 109.0 &	      & 	 &	     \\
  01:05  &  01:10 &	      &  22.73  & 108.5 &	      & 	 &	     \\
  01:13  &  01:18 &	      &  23.25  & 105.7 &	      & 	 &	     \\
\tableline
\end{tabular}
}
\end{center}
\end{table*}

\clearpage

\begin{table*}
\begin{center}
\caption{Results of the modeling of the CHARA observations.\label{tab:CHARAresults}. {\bf $N$ is the
number of data points.}}
\begin{tabular}{lrr}
\tableline\tableline
Quantity & Value & Estimated uncertainty \\
\tableline
Surface brightness ratio in $K_s$ band     & $0.330$     & $\pm 0.01$\\
{\bf Ascending node}                       & $48\degr$ & $\pm 2\degr$\\
Angular size of the semimajor-axis (mas)   & $2.28$      & $\pm 0.02$\\
{\bf $\chi^2/(N-1)$}                       & $3.76$      & \\
\tableline
\end{tabular}
\end{center}
\end{table*}

\clearpage

\begin{table*}
\begin{center}
\caption{Calculated orbital elements of Algol B from EVN data. (The first {\bf line} at $\Omega_1$ and $\chi^2$ belongs to the 'fixed $a$' solution, while 
the second one to the 'adjusted $a$' results. Errors are $1\sigma$ errors.)
\label{tab:kettospalya}}
\begin{tabular}{lcrr}
\tableline\tableline
Quantity & Notation & Value & Formal error \\
\tableline
Period                  & $P_1$      & $2\fd8673$               & fixed   \\
Semi-major axis         & $a_B$    & $0\farcs0019$            & fixed   \\
                        &          & $0\farcs00353$           & $0\farcs00005$   \\      
Eccentricity            & $e_1$      & 0                        & fixed	\\
Inclination             & $i_1$      & $82\fdg3$                & fixed   \\
Argument of periastron  & $\omega_1$ & 0\tablenotemark{a}       & fixed   \\
Longitude of the ascending node & $\Omega_1$ & $53\degr$        & $826\degr$\\
                        &          & $53\degr$                & $239\degr$\\
 {\hbox{   \it from linear fit}}&  & $52\degr$                & $3\degr$ \\
Mean anomaly at $t_0$   & $(M_0)_1$    & $90\degr$                & fixed   \\
\tableline
Epoch                   & $t_0$    & $2\,454\,084.360$\tablenotemark{b} & $-$ \\
\tableline
                        & $\chi^2$ & $0.076623844$            & \\
                        &          & $0.025291438$            & \\
\tableline\tableline
\multicolumn{4}{l}{Quantities for the corrections} \\
\tableline
Trigonometric parallax  & $\pi$    & $0\farcs035$              & \\
Proper motion components& $\mu_\alpha$ & $0\fs031$cent$^{-1}$    & \\
                        & $\mu_\delta$ & $-0\farcs09$cent$^{-1}$ & \\
\tableline
\multicolumn{4}{l}{Orbital elements of Algol AB in ternary system\tablenotemark{c}} \\
\tableline
Period                  & $P_2$      & $679\fd276349353$        & \\
Semi-major axis         & $a_{AB}$      & $0\farcs025738139$       & \\
Eccentricity            & $e_2$      & $0.212719923$            & \\
Inclination             & $i_2$      & $84\fdg014938082$        & \\
Argument of periastron  & $\omega_{AB}$ & $132\fdg558322883$       & \\
Longitude of the ascending node & $\Omega_2$ & $312\fdg345789012$  & \\
Time of periastron      & $T_2$      & $2\,446\,937.879685247$  & \\
\tableline

\end{tabular}
\tablenotetext{a}{In case of circular orbit $\omega$ is undetermined. It was formally set to zero.
This means that the mean anomaly ($M$) is measured from the ascending 
node.}
\tablenotetext{b}{Mid-eclipse moment of secondary minimum occurred during the EVN observation.}
\tablenotetext{c}{Adopted from our recalculations of Pan et al. (1993) measurements.}
\end{center}

\end{table*}

\clearpage

%
%

\begin{figure} 
\epsscale{.50}
\includegraphics[angle=0,scale=.50]{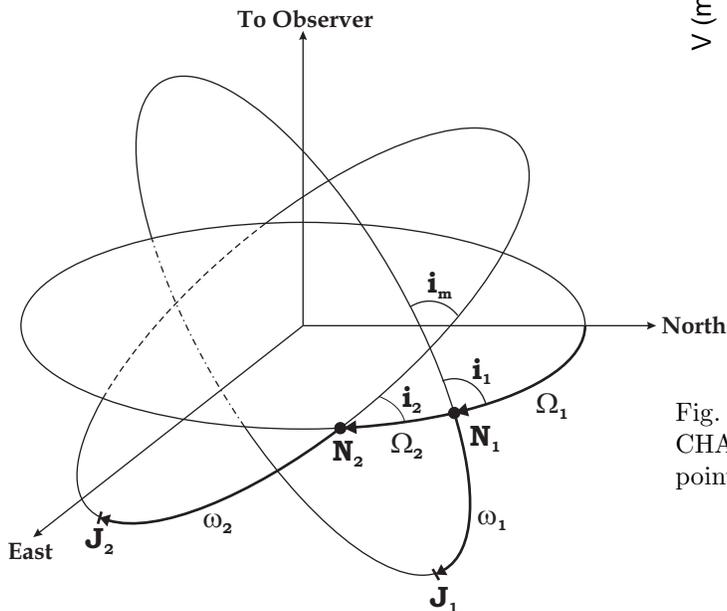} 
\caption{The meaning of different angular orbital elements mentioned in this 
paper. Orbits
are projected into a sphere. $\Omega_1$, $\Omega_2$ are the longitudes 
of the
nodes of the close and the wide pair, respectively, $N_1$, $N_2$ are the
ascending nodes. $i_1$ and $i_2$ denotes the inclinations of the orbits
measurable by an  observer while $i_m$ is the mutual inclination of the two
orbital planes. $J_1$ and $J_2$ are the pericentre points. \label{fig:Algol_coordinate_system}}
\end{figure}

\begin{figure}
\epsscale{.50}
\includegraphics[angle=270,scale=.50]{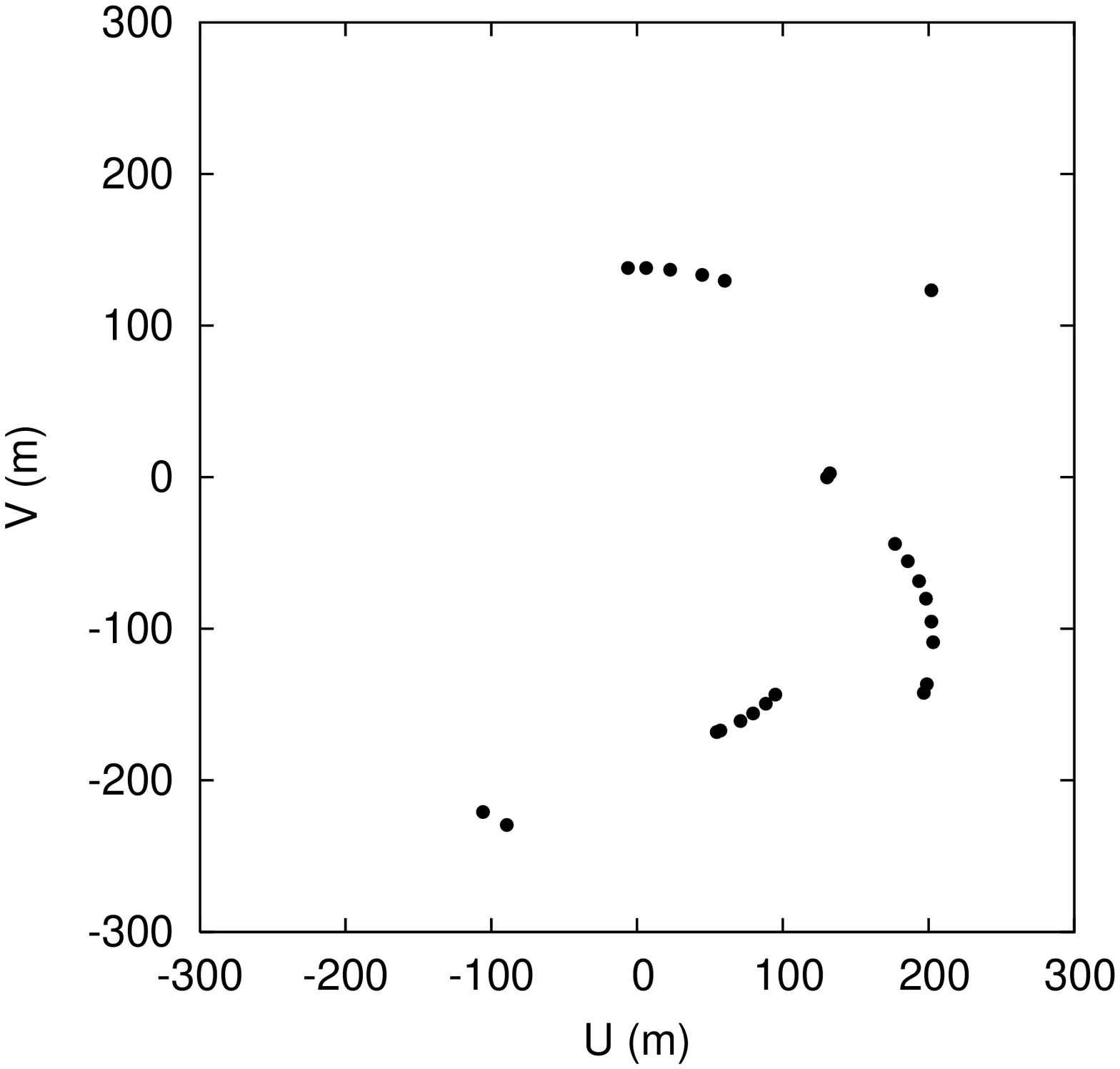}
\caption{The UV coverage of the Algol with CHARA. Each point represents one observational
point.\label{fig:UVCoverage}}
\end{figure}

\begin{figure}
\epsscale{.50}
\includegraphics[angle=0,scale=.50]{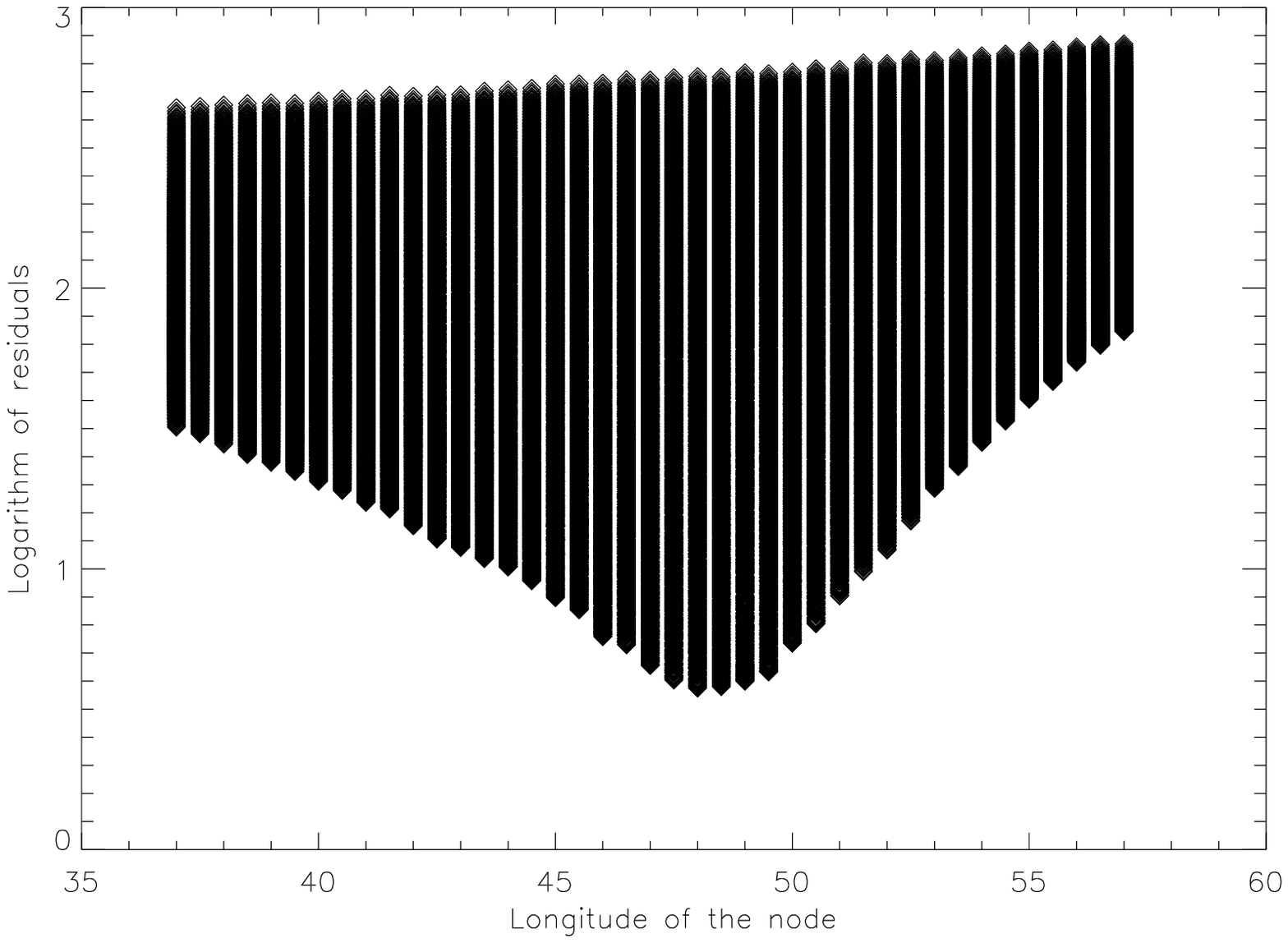}
\caption{The result of the grid search for minimum value of $\chi^2$.
Only the result we obtained for $\Omega_1$ is shown here. Note that the {\bf decimal} logarithm
of the {\bf $\chi^2/(N-1)$ (N is the number of the data points)} is 
shown for the sake of a better visualization.\label{fig:Omegamin}}
\end{figure}

\begin{figure}
\epsscale{.50}
\includegraphics[angle=0,scale=.50]{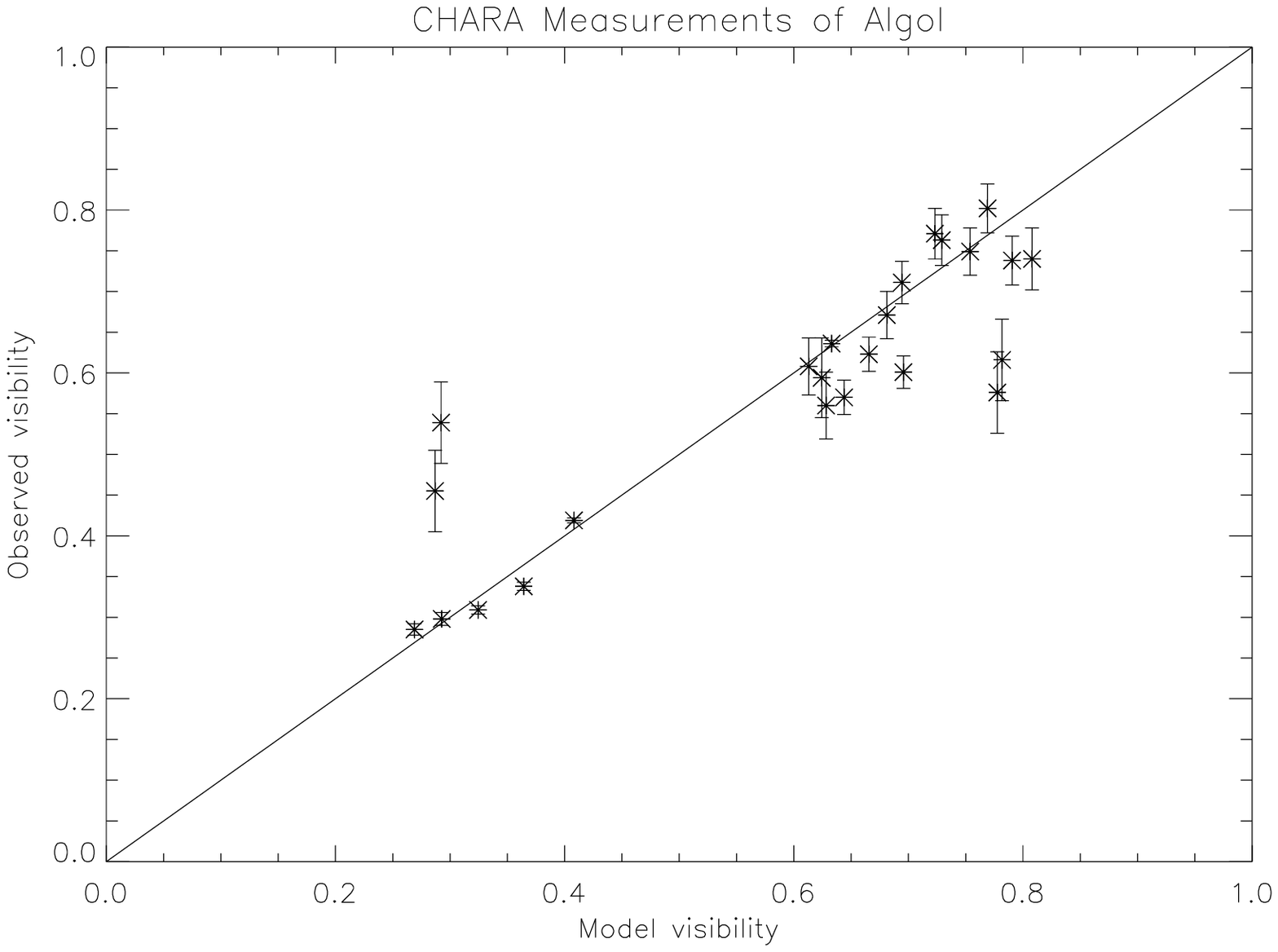}
\caption{The modelled versus observed visibility values and their errors. The
solid line shows the 1:1 line.\label{fig:CHARAFit}}
\end{figure}

\begin{figure}
\epsscale{.40}
\includegraphics[angle=0,scale=.40,bb=36 120 576 620,clip]{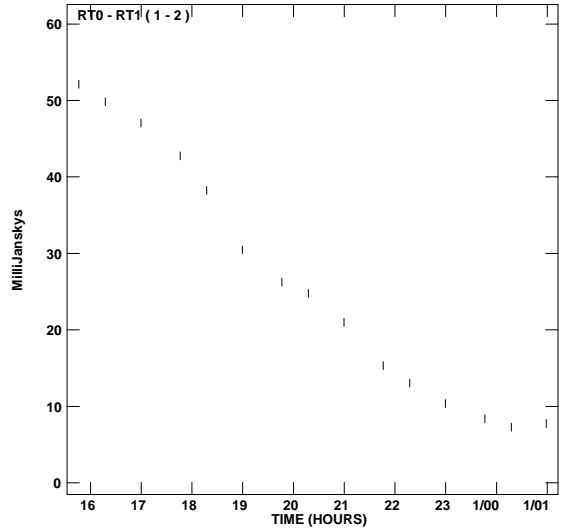}
\caption{Radio total flux density variation of Algol from the Westerbork 
synthesis array data, taken during the VLBI observations. At the start of 
the observations there was a radio flare.
\label{fig:flare}}
\end{figure}

\begin{figure}
\epsscale{.70}
\includegraphics[scale=.70]{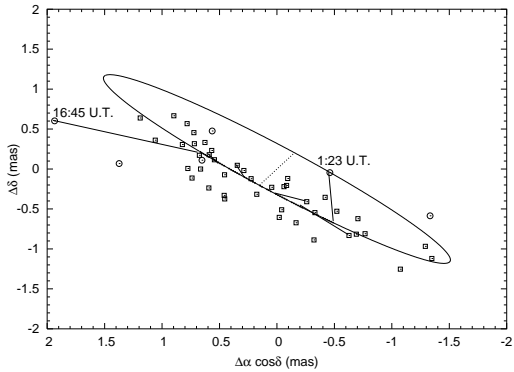}
\caption{\bf The 5-minutes averaged data points with our astrometric fit 
of component B (see Table~\ref{tab:kettospalya} for the
orbital elements). 
The circled points were excluded from the fittings due to their large 
scatter. We labeled our first
(16:45 heliocentric UT) and last (1:23 heliocentric UT) points. Furthermore, the $10^{th}$, $20^{th}$, $30^{th}$
and $40^{th}$ points are also connected to their theoretical positions on the orbit. The slim dashed
line represents the line which connects the mid-eclipse points of the 
primary and the secondary eclipses. 
\label{fig:kettospalya}}
\end{figure}

\begin{figure}
\epsscale{.70}
\includegraphics[scale=.70]{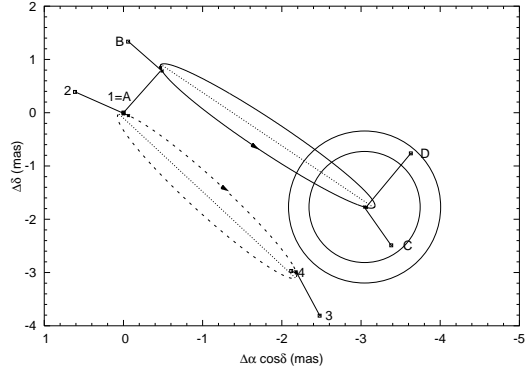}
\caption{\bf The astrometric orbit of Algol B for the four positions measured by Lestrade et al. (1993).  Points
1 to 4 represents the points after Lestrade et al.'s original corrections, while points A to D
denote our recalculated positions (see text for details). (Note that the absolute coordinates of
points 1 and A are not equal, but in this figure relative coordinates are used.) The observed points
are connected by short lines with their corresponding positions along the astrometric orbits. The
astrometric orbits are also plotted: dashed lines represents Lestrade et al.'s orbit while solid
line corresponds to our recalculated orbit. The slim dashed lines represent the nodal line.  (The
ascending nodes are in the vicinity of points 2 and B, respectively). The arrows along the orbits
show the direction of the orbital revolution. (Note, that the original solution of  Lestrade et
al.~1993 gives a reverse direction.) The arrows are located in the mid-eclipse point of the
secondary minimum, where the mean anomly is equal to $90\degr$. The inner of the two circles,
centered at the corresponding point along the recalculated orbit of position C represents the
surface of the Algol B component, while the outer one illustrates 
schematically the separation of
the two radio emitting lobes detected by Mutel et al.~(1998).
\label{fig:Lestrade}}
\end{figure}

\begin{figure}
\epsscale{.70}
\includegraphics[scale=.70]{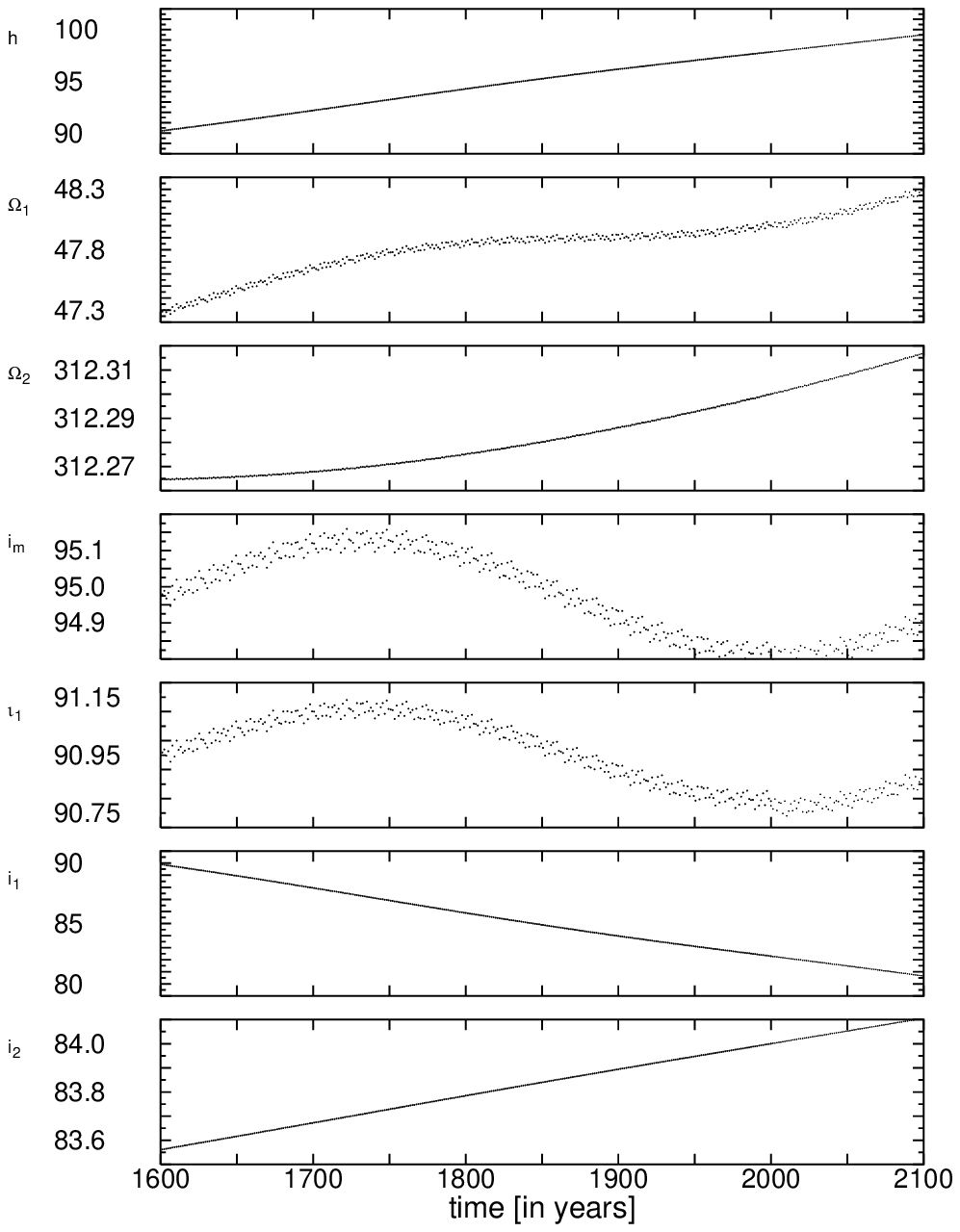}
\caption{The variation of the angular orbital elements of Algol system between AD 1600 and 2100.
The panels from top to bottom: $(h)$ node of the binary measured 
in the invariable plane measured from the intersection of the invariable plane and the sky;
($\Omega_1$) node of the close binary; ($\Omega_2$) same for the third component;
($i_\mathrm{m}$) mutual inclination; ($\iota_1$) inclination of the binary with 
respect to the invariable plane;
($i_1$) inclination of the close binary; ($i_2$) same for the tertiary. 
 \label{fig:Algol_500years}}
\end{figure}

\begin{figure*}
\epsscale{.70}
\includegraphics[scale=.70]{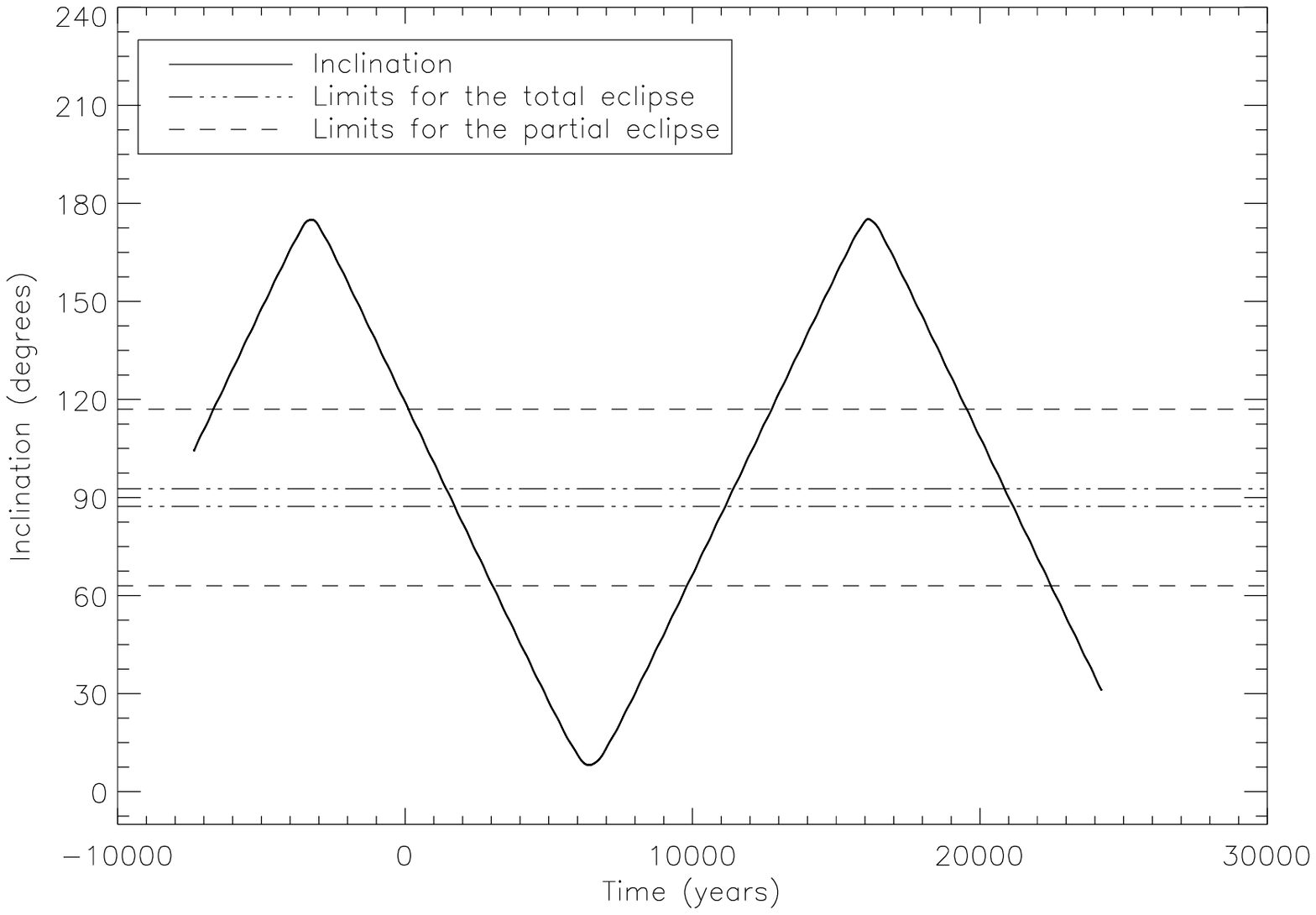}
\caption{The variation of the inclination of Algol AB with time. Solid line
represents the inclinitaion of the close pair observable from the Earth.
Between the dashed lines the system shows partial eclipses, moreover between
the dashed-dotted lines the system shows total eclipses for a short time.
 \label{fig:Algol_long}}
\end{figure*}

\end{document}